\documentclass[12pt]{article}
\usepackage{epsfig}

\parskip=\medskipamount % space between paragraphs (TeX)
plus.3em minus.5em \abovedisplayshortskip=.5em plus.2em minus.4em
\renewcommand{\thesection}{\arabic{section}}

\catcode`@=11

\@addtoreset{equation}{section} \@addtoreset{equation}{subsection}
\def\theequation{\ifnum\value{section}=0 \arabic{equation}\ignorespaces
\else \ifnum\value{section}=-1 A.\arabic{equation}\ignorespaces
\else \ifnum\value{subsection}=0
\thesection.\arabic{equation}\ignorespaces \else
\thesection.\arabic{subsection}.\arabic{equation}\ignorespaces
                             \fi
                        \fi
                   \fi}

{\catcode`\'=\active \def'{{}^\bgroup\prim@s}}

\catcode`@=12

\newcommand{\bq}{\begin{equation}}
\newcommand{\be}{\begin{equation}}
\newcommand{\fq}{\end{equation}}
\newcommand{\ee}{\end{equation}}
\newcommand{\bqr}{\begin{eqnarray}}
\newcommand{\beqs}{\begin{eqnarray}}
\newcommand{\fqr}{\end{eqnarray}}
\newcommand{\eeqs}{\end{eqnarray}}

\newcommand{\rf}[1]{(\ref{#1})}

\def\bop#1{\setbox0=\hbox{$#1M$}\mkern1.5mu
    \vbox{\hrule height0pt depth.04\ht0
    \hbox{\vrule width.04\ht0 height.9\ht0 \kern.9\ht0
    \vrule width.04\ht0}\hrule height.04\ht0}\mkern1.5mu}
\begin{document}
\thispagestyle{empty}

\vskip .6in
\begin{center}

{\bf Possible Solution to the Poincare Conjecture}

\vskip .6in

{\bf Gordon Chalmers}
\\[5mm]
% {\em address \\
%      address \\
% Los Angeles, CA } \\

{e-mail: gordon@quartz.shango.com}

\vskip .5in minus .2in

{\bf Abstract}

\end{center}

The Poincare conjecture is analyzed in the context of 
Calabi-Yau $n$-folds.  A simple treatment is given by embedding the 
three-manifolds into these CY manifolds, and then taking the orbifold 
limit.  The higher-dimensional proofs are also available in this context. 

\vfill\break

The Poincare conjecture states essentially that a simply connected 
three-manifold is homeomorphic to the $S^3$.  The proofs in four and 
higher dimensions are contained in \cite{Freedman} and \cite{Smale}.  
The analagous statement for non-simply connected three manifolds is 
that they are homeomorphic to the $S^3/\Gamma$, with $\Gamma$ a 
discrete subgroup of SO(4).  

A possible proof here is proferred by embedding the metric components of the 
$3$-manifold into a Ricci-flat Calabi-Yau manifold.  The Calabi-Yau manifolds 
have orbifold limits into $T^{2n}/Z_p$; as such the metric components become 
functions of the orbifold in the smooth degeneration of the Calabi-Yau into 
the orbifold.  The curvature of the metric solution on the tori are constant, 
hence models a sphere.  

Consider the $6$ metric components of a Reimannian 3-manifold, $g_{\mu\nu}$.  
Take these metric components as holomorphic coordinates of a Calabi-Yau $
6$-fold, $X=g$.  The copy of the metric coordinates models the 
anti-holomorphic coordinates, ${\bar X}$.  The set of points generated 
by the $3$-manifold's metrics then span a locus of points in the Calabi-Yau 
$6$-fold, possibly 
the entire holomorphic half of the $6$-fold.  The curvature of the metric 
on the space of $g$ on a $T^3/\Gamma$ has curvature zero, and the blow up 
is found by a diffeomorphism effectively which 
does not change the curvature condition.  Several patches are required 
to avoid unphysical singularities, with zero curvature (one reason 
is by self-consistency as the space is homeomorphic to a $T^3/\Gamma$).   
The embedding is chosen by using the full metric (periodic) basis of 
the $3$-manifold.  In the case of manifolds with higher homotopy, the 
singularity of the $3$ manifold is placed at the center of one of the 
CY $2$-spheres in order to coincide with the orbifold limit of 
$S^3/\Gamma$, i.e. all the $2$-spheres line up. 

The curvature condition of the individual Calabi-Yau manifolds is $R=0$.  
However, since the coordinates of these manifolds are the metric, the 
translation of this curvature into the $3$-manifold's curvature is 
complicated, 

\bqr 
R(y)=\Lambda(y) \ , 
\label{curvature}
\fqr  
with $\Lambda(y)$ a non-negative function.  This can be seen to be 
non-negative after the fact of taking the orbifold limit of the Calabi-Yau.   
A diffeomorpism of the $3$-manifold is required to map the non-negative 
curvature into the form $\Lambda(y)$, together with the appropriate 
choice of the Calabi-Yau $6$-fold.  

The Calabi-Yau manifolds have orbifold limits of the form, $M_6\rightarrow 
T^{12}/Z_p$, holomorphically.  As the metric components $g$ of the 
$3$-manifold 
are the coordinates of the Calabi-Yau, in the smooth orbifold limit they 
span a locus of coordinates of the $T^{12}/Z_p$.  The toriodal limit 
produces the individual curvatures of the metric components as if they 
were on circles, and they follow as coordinates on $Z(y)=Z(y+2\pi R)$.  

The constant circle curvatures of these deformed metric components model 
that of 
a $3$-manifold with constant curvature in all directions in the metric 
components.  The question is how this is related to a $3$-sphere.  The 
holomorphic space parametrized by such metric components $X=g$ has the 
topology of a $T^3/Z_q$, with zero curvature, but with constant curvature 
which is non-vanishing 
along the circles.   This has the same topology as a $3$-sphere.  
Furthermore,  the homotopy of the original $3$-manifold is trivial, 
this requires the coordinates resulting from metric components, to be 
also trivial.    

The original curvature condition of the $3$-manifold follows from 
\rf{curvature}.  
A diffeomorphism of the $3$ manifold can change this non-negative curvature 
into other functions $\tilde\Lambda(y)$.  Or rather in the converse, what is 
required is to change the original $3$-manifold's curvature into the form 
of \rf{curvature}.  This curvature condition depends on the specific 
Calabi-Yau, in finding the conditions $R_6=0$ and $R_3(y)=\Lambda(y)$.    
  
Thus holomorphically embedding the components of the $3$-manifold into the 
$6$-fold, together with the known smooth degenerations of the latter, generate 
the homeomorphisms of the former into the $3$-spheres.  The proof can 
also be run backwards, by starting with a $3$-sphere and blowing it up 
into the Calabi-Yau, after including its anti-holomorphic half.

The same argument goes through for higher dimensional simply connected 
manifolds.  Furthermore, the argument seems to show how Thurston's 
conjecture holds, for non-simply connected $3$-manifolds and $S^3/\Gamma$.

\vskip .2in 
Gordon Chalmers thanks Albert Whitehead of the Royal Theological Society 
of Newark.

\vfill\break

\end{document}